# A Robust Synthesis of Fluorosurfactants with Tunable Functions via a Two-Step Reaction

Jiyuan Yao[a,c], Shijian Huang[b], Shuting Xie[a], Zhenping Liu[a], Yueming Deng[a], Luca Carnevale[c], Mingliang Jin[a], Loes I. Segerink[c], Da Wang[d*], Lingling Shui[a,b*], and Sergii Pud[c]

**a.** International Joint Laboratory of Optofluidic Technology and System (LOTS), National Center for International Research on Green Optoelectronics, Guangdong Provincial Key Laboratory of Nanophotonic Functional Materials and Devices, South China Academy of Advanced Optoelectronics, South China Normal University, Guangzhou, 510006 (China)

**b.** School of Optoelectronic Science and Engineering, South China Normal University, Guangzhou, 510006 (China)

**c.** BIOS Lab-on-a-chip group, EEMCS Faculty, MESA+ institute, University of Twente, Enschede, 7500 AE (The Netherlands)

**d.** South China Academy of Advanced Optoelectronics, South China Normal University, Guangzhou, 510006 (China)

* email address: shuill@m.scnu.edu.cn; da.wang@m.scnu.edu.cn

# Abstract:

Fluorosurfactant-stabilized microdroplets hold significant promise for a wide range of applications, owing to their biological and chemical inertness. However, conventional synthetic routes for fluorosurfactants typically require multiple reaction steps and stringent conditions, such as high temperatures and anaerobic environments. This complexity poses a significant limitation to the development of fluorosurfactant synthesis and their subsequent applications in droplet-based systems. In this work, we present a robust two-step synthesis of fluorosurfactants with tunable functionalities. Microdroplets stabilized by these fluorosurfactants exhibit enhanced stability and biocompatibility. Notably, these fluorosurfactants facilitate the formation of nanodroplets that efficiently transport and concentrate fluorophores with high selectivity. Furthermore, we demonstrate that colloidal self-assemblies with tunable morphologies can be engineered by modulating interactions between the fluorosurfactants and colloidal particles. Our synthetic approach provides a strategy for the rapid production of functional fluorosurfactants under mild conditions, enabling droplet-based microfluidic techniques with applications in biology and material science.

# 1. Introduction

Droplet microfluidics stands as a versatile technology for micrometer-scale total analysis systems (μTAS) and advancing science across diverse research fields. It enables the generation and manipulation of isolated droplets with volumes ranging from picoliters to nanoliters on micro-chip[1,2]. Due to their excellent gas permeability and compatibility with polydimethylsiloxane (PDMS) and poly(methyl methacrylate) (PMMA)-based microfluidic chips, fluorinated oil are commonly applied as continuous phase in generating water-in-fluorinated oil (W/O$_F$) microdroplets with adjustable diameters and components. The generated microdroplets have broad applications, including the construction of spheroids and organoids[3,4,5], single cell analysis[6,7] and the structuring self-assemblies with tunable functions[8,9].

Surfactants play a crucial role in both the emulsification and stabilization of emulsions in droplet microfluidics[10,11]. In contrast to conventional surfactants (*e.g.*, Span 80, Abil EM 90, Abil EM 180), fluorosurfactants, consisting of hydrophobic and oleophobic "fluorinated tail" and hydrophilic "head" groups, exhibit superior compatibility with fluorinated oil. This enhances their ability to prevent coalescence in water-in-fluorinated oil (W/O$_F$) or organic solvent-in-fluorinated oil microdroplets[8,12,13]. Over the past decades, a variety of fluorosurfactants with different hydrophilic groups have been developed[14,15,16,17,18]. A common strategy for synthesizing fluorosurfactants involves leveraging electrostatic interactions between polyetherdiamine additives in the aqueous phase and carboxylated perfluorocarbon surfactants in the fluorinated oil phase[19]. However, this method often compromised the stability of fluorosurfactants. For example, polycationic biomarkers containing amine groups can be adsorbed at the droplet interface through weak electrostatic interactions, interfering with the formation of stable fluorosurfactants and affecting their performance in bioimaging and screening applications[14,20]. To address these limitations, covalently linking hydrophilic segments with fluorinated tails has emerged as an effective approach, leading to commercially available products such as PFPE(H)$_2$-PEG$_{600}$ (RAN Biotechnologies) and Pico-Surf$^{TM}$ (Dolomite Microfluidics, Royston, UK)[18]. However, the complex synthetic procedures — such as amide linkage[14], click chemistry[21,22], and living radical polymerization[18], — along with the use of moisture-sensitive reagents (*e.g.*, thionyl chloride and perfluoropolyether-based (PFPE) fragmentation chain transfer (RAFT) reagent), constrain their application in droplet microfluidics. Furthermore, the synthesis of customized fluorosurfactants with tunable functions, such as stimulus response[23,24], or interfacial interactions[25,26], remains largely unexplored. Therefore, there is a pressing need for a simpler, more efficient synthetic route for functional fluorosurfactants to advance droplet-templating microfluidic techniques.

In this work, we introduce a two-step synthetic strategy to prepare fluorosurfactants with tunable functionalities. Our approach begins with the synthesis of a customized fluorosurfactant PFPE(H)-COOCOOC$_2$H$_5$, *via* a mixed anhydride reaction, followed by the amidation-based synthesis of commercially available fluorosurfactants (PFPE(H)$_2$-ED$_{900}$ and PFPE(H)$_2$-PEG$_{600}$) through amidation reaction. This method offers several advantages over previous protocols, including reduced reaction time and milder reaction conditions. We demonstrate that these fluorosurfactants enhance microdroplet stability across a broad temperature range and exhibit excellent biocompatibility, facilitating applications such as yeast culturing. Moreover, we demonstrate that these fluorosurfactants enable tunable functionalities: 1) nanodroplets, formed *via* spontaneous emulsification, selectively transport and enrich fluorophores, and 2) two distinct superstructures were obtained through the self-assembly of nanoparticles (NPs) from drying droplets, driven by tunable interactions between the fluorosurfactants and NPs. Our synthesis strategy expands the toolbox for developing functional fluorosurfactants and opens new avenues for the design of fluorinated oil-based droplet microfluidics with tunable properties.

## 2. Results and Discussion

## 2.1 Two-step reaction synthesis of fluorosurfactants

We synthesized a diblock fluorosurfactant (PFPE(H)-COOCOOC$_2$H$_5$) through a mixed anhydride reaction, coupling fluorinated tail of carboxylated perfluoropolyether (PFPE-COOH) and ethyl chloroformate at 0 ºC for 0.5 h. Subsequently, triblock fluorosurfactant (PFPE(H)$_2$-ED$_{900}$) was obtained *via* amidation reaction, coupling the PFPE(H)-COOCOOC$_2$H$_5$ with hydrophilic segment Jeffamine ED$_{900}$ (NH$_2$-ED$_{900}$-NH$_2$) at RT for 0.5 h (Fig. 1a; see Methods for details). Fourier-transform infrared spectroscopy (FT-IR) (Fig. 1b) and $^{19}$F-nuclear magnetic resonance ($^{19}$F-NMR) spectroscopy (Fig. 1c; Supplementary Fig. 1) were employed to characterize the chemical nature of the PFPE(H)-COOH, PFPE(H)-COOCOOC$_2$H$_5$ and PFPE(H)$_2$-ED$_{900}$, respectively. It is noted that carboxylic group (C=O) stretch at 1774 cm$^{-1}$ in PFPE(H)-COOH red-shifted[16], and two new absorbance peaks at 1795 cm$^{-1}$ and 1785 cm$^{-1}$ in PFPE(H)-COOCOOC$_2$H$_5$ emerged, corresponding to the C=O stretch of the converted mixed anhydride. Upon coupling PFPE(H)-COOCOOC$_2$H$_5$ with Jeffamine ED$_{900}$, the C=O stretch at 1795 cm$^{-1}$ and 1785 cm$^{-1}$ disappeared, and a new peak at 1693 cm$^{-1}$ emerged, corresponding to the amide C=O stretch. The peak of $^{19}$F-NMR signal at -133.25 ppm, appearing as a quartet, corresponds to the fluorine absorption vibration on the adjacent PFPE carboxyl carbon atoms[24]. When ethyl chloroformate was covalently bound to the carboxyl group of PFPE(H)-COOH, this peak shifted to -133.00 ppm. The peak disappeared upon coupling with the amino group of Jeffamine ED$_{900}$ (NH$_2$-ED$_{900}$-NH$_2$), confirming the formation of amide linkage in PFPE(H)$_2$-ED$_{900}$. Both FT-IR and $^{19}$F-NMR spectroscopy results consistently confirm the successful synthesis of the fluorosurfactants[14, 24].

The robustness of our proposed synthetic strategy at room temperature (RT) enables the incorporation of various hydrophilic segments with thermo-sensitivity. We further demonstrated the successful synthesis of triblock fluorosurfactant (PFPE(H)$_2$-PEG$_{600}$), using a PEG-diamine hydrophilic segment, achieving a yield approximately 90% (Supplementary Fig. 2a-c). In comparison to previously reported protocols, we remark that our fluorosurfactants can be obtained under a mild condition and atmospheric environment within a relatively short reaction time. These advantages underscore the significant potential of our method for industrial-scale production[14,16,17,24].

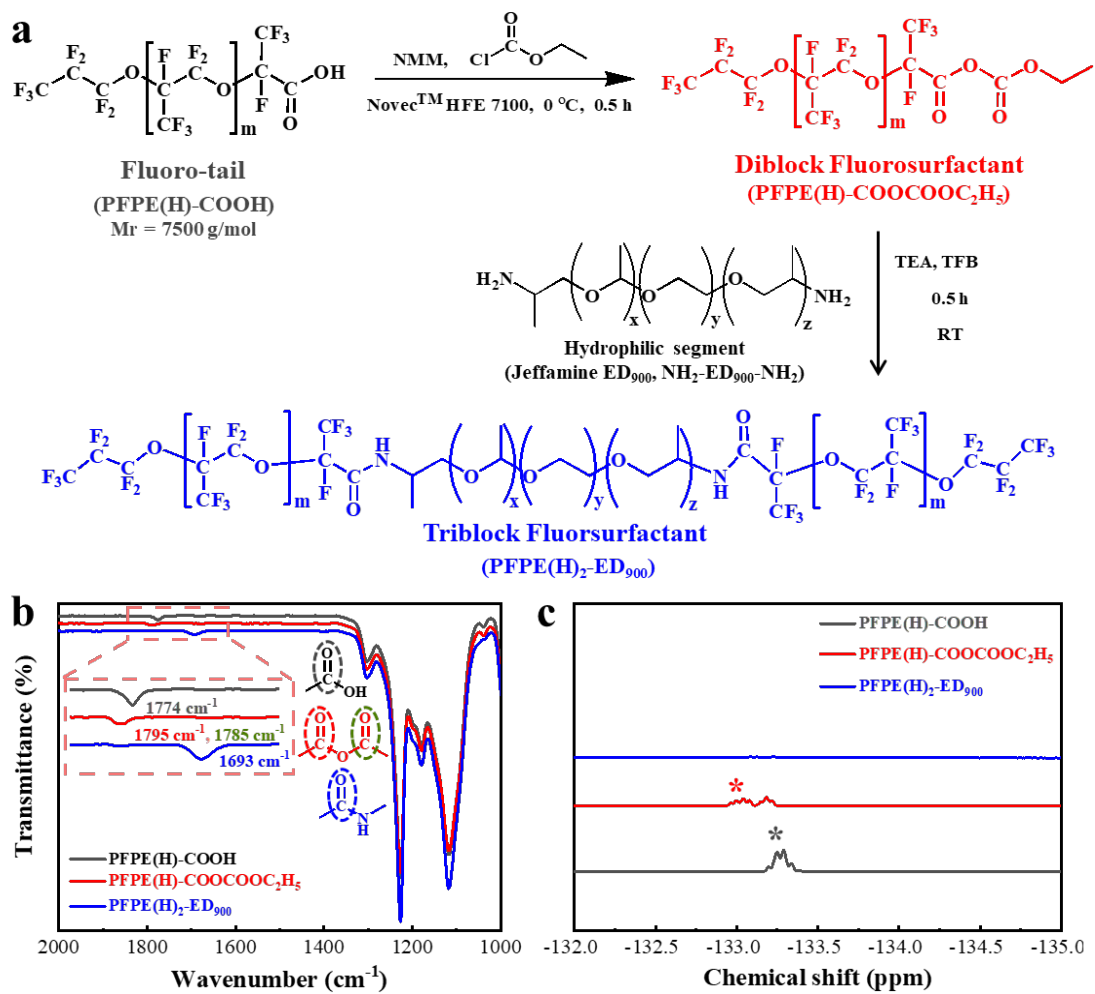

**Figure 1. Synthesis and characterization of the fluorosurfactants.** a) Schematic illustration of synthetic procedure for the fluorosurfactants in this study. b) FT-IR spectra and c) $^{19}$F NMR spectra of PFPE(H)-COOH, PFPE(H)-COOCOOC$_2$H$_5$ and PFPE(H)$_2$-ED$_{900}$ synthesized at RT for 0.5 h. Inset of panel b highlights the shift of the C=O stretch during the synthesis of the fluorosurfactants. Note that the asteroid in panel c denotes fluorine absorption vibration on the adjacent PFPE(H)-COOH carboxyl carbon atom in these three fluorosurfactants.

## 2.2 Stability and biocompatibility of W/O$_F$ microdroplets using the synthesized fluorosurfactants

To investigate surface activity of our fluorosurfactants in W/O$_F$ droplets, we measured interfacial tension γ between deionized water (DI water) and Novec™ HFE 7500 oil *via* pendant drop experiments (Supplementary Fig. 3). The interfacial tension dropped as the concentration of the three fluorosurfactants increased. It was observed that the interfacial tension was reduced from ~45.43 mN/m to ~29.64 mN/m, ~19.90 mN/m and ~16.83 mN/m when PFPE(H)-COOCOOC$_2$H$_5$, PFPE(H)$_2$-ED$_{900}$ and PFPE(H)$_2$-PED$_{600}$ were applied, respectively, indicating that the synthesized fluorosurfactants exhibit good surface activity. We further calculated the critical micelles concentration (CMC) of the three fluorosurfactants, as summarized in Supplementary Table 1.

Water-soluble fluorophores are ideal for tracking droplet stability, as changes in fluorescence intensity and droplet size can be directly visualized, allowing straightforward monitoring of droplet integrity. To evaluate the droplet-stabilizing capability of the synthesized fluorosurfactants, we prepared (W/O$_F$) microdroplets containing FITC-CM-Dextran at the concentration of 3.0 mg/mL in Novec™ HFE 7500 oil, stabilized with different fluorosurfactants at a concentration of 5.0 mM (above their CMC). This was achieved using a flow-focusing PDMS microfluidic device (Supplementary Fig. 4a; Supplementary Fig. 5a-b). The generated microdroplets were collected into a home-made poly (methyl methacrylate) (PMMA) cell and continuously incubated at RT (Supplementary Fig. 5c-d).

The size of the microdroplets stabilized with PFPE(H)-COOCOOC$_2$H$_5$ remained largely unchanged after 5-day incubation (upper panel of Fig. 2a), and the corresponding fluorescence intensity of the microdroplets decreased by approximately 5.2 % during the first day of incubation and then remained nearly unchanged (flotation within ± 1.4 %) over the following days (bottom panel of Fig. 2a), indicating excellent stability. Similar stability in size of microdroplet and the corresponding fluorescence intensity (flotation within ± 2.9 %) was observed for microdroplets stabilized with both PFPE(H)$_2$-ED$_{900}$ and PFPE(H)$_2$-PEG$_{600}$ at RT (Supplementary Fig. 6a-b). Notably, our fluorosurfactants also preserved the stability of low molecular weight fluorophores such as rhodamine 110 chloride and resorufin sodium salt under ambient conditions (Supplementary Fig. 7a-f). It is worth noting that there was no phase separation or coalescence of the droplets during storage at RT. In contrast, we observed the fluorescence intensity decayed in commercial fluorosurfactants (*e.g.*, Pico-Surf™) stabilized microdroplets within two-day incubation, likely due to leakage or bleach of resorufin sodium salt (Supplementary Fig. 8). This comparison demonstrates the superior droplet-stabilizing properties of our synthesized fluorosurfactants — PFPE(H)$_2$-ED$_{900}$ and PFPE(H)$_2$-PEG$_{600}$ — by preventing loss of the fluorophores under ambient conditions. Moreover, we observed that microdroplets stabilized by PFPE(H)$_2$-ED$_{900}$ and PFPE(H)$_2$-PEG$_{600}$ retained their integrity and size after multiple thermal treatments up to 95 ºC (Supplementary Fig. 9a-b), highlighting the potential of these fluorosurfactants in biochemical applications, such as polymerase chain reaction (PCR)[18] and screening assays[27].

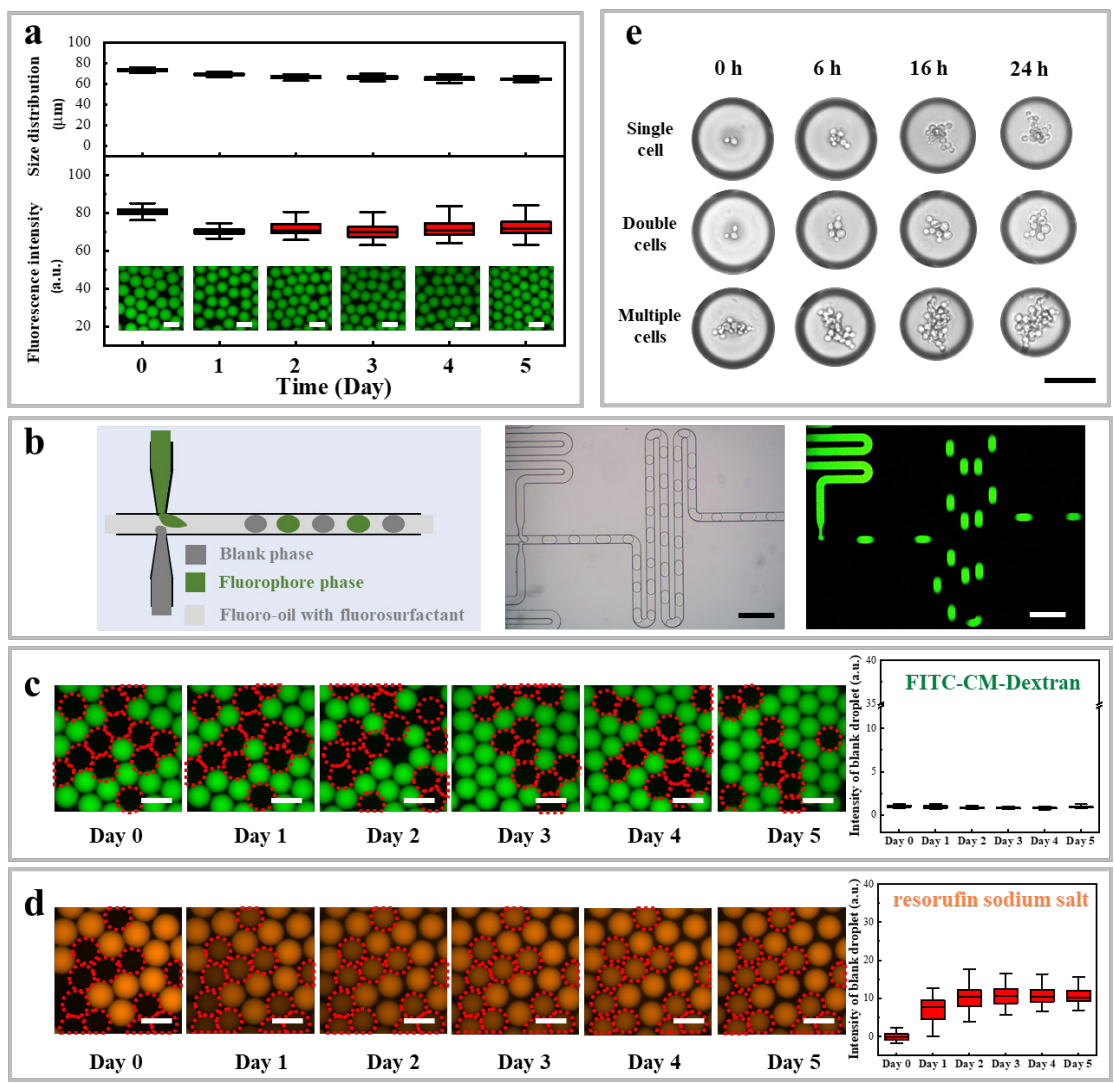

**Figure 2. Stability and biocompatibility of microdroplets stabilized by the synthesized fluorosurfactants.** a) Size distribution and fluorescence intensity of FITC-CM-Dextran in W/O$_F$

microdroplets stabilized with PFPE(H)-COOCOOC$_2$H$_5$ at the concentration of 5.0 mM over a five-day stability test at RT, error bars representing 1.5 IQR. b) Schematic illustration (left), bright-field micrograph (middle), and fluorescence micrograph (right) of microdroplets with and without fluorophores generated by a double T-junction generator. c,d) Fluorescence microscopy images, along with a bar chart of fluorescence intensity of blank W/O$_F$ microdroplets over five days, error bars representing 1.5 IQR. e) Yeast cell culturing in W/O$_F$ microdroplets stabilized with PFPE(H)-COOCOOC$_2$H$_5$, demonstrating the growth of single cell, double cells and multi-cells within individual W/O$_F$ microdroplets. All experiments in this figure used PFPE(H)-COOCOOC$_2$H$_5$ at the concentration of 5.0 mM at RT. Note that the fluorescence intensity was extracted from the raw (gray scale) fluorescence micrographs, with false colors applied to enhance visibility of the distribution and intensity of the fluorophores. Scale bars: 100 μm (a,c,d), 500 μm (b) and 50 μm (e).

However, these retention measurements of fluorophore cannot eliminate the role of reverse-micelles-mediated mass transport between adjacent microdroplets, which may result from fluorophore diffusion driven by osmotic pressure differences arising from concentration gradients. To assess the performance of our fluorosurfactants in preventing inter-droplet mass transfer, we further examined the transfer of fluorophores from fluorosurfactants-stabilized W/O$_F$ microdroplets to blank W/O$_F$ microdroplets (without fluorophores) (Fig. 2b; Supplementary Fig. 4b and Methods section for droplet generation details). Remarkably, regardless of the fluorosurfactant applied, almost no diffusion of FITC-CM-Dextran was observed from the loaded microdroplets to the blank microdroplets over a five-day period (Fig. 2c; Supplementary Fig. 10). Both loaded and blank microdroplets retained their original sizes, with no coalescence detected. This behavior can be attributed to the chemically inert nature of the fluorosurfactants, which effectively mitigated inter-droplet mass transfer, and the high molecular weight of FITC-CM-Dextran.

To further understand the effect of the properties of the fluorophores on the inter-droplet mass transfer, we studied the transfer of resorufin sodium salt. Despite of minor transfer of resorufin sodium salt from loaded microdroplets to the blank ones on Day 1, the fluorescence intensity remained almost identical from Day 1 to Day 5 (Fig. 2d; Supplementary Fig. 11a-b). We attribute the inter-droplet transfer behavior to the higher water solubility of resorufin sodium salt and lower molecular weight compared to FITC-CM-Dextran. It is important to note that the fluorosurfactants were applied at a concentration of 5.0 mM (above the critical micelle concentration, CMC), a condition under which reverse micelles (RMs) are expected to form. As a result, the self-assembled RMs from the fluorosurfactants most likely solubilized low-molecular-weight fluorophore molecules (*e.g.*, resorufin sodium salt in this study), which may partially explain the observed minor inter-droplet transfer.

Beyond their effectiveness in inhibiting inter-droplet mass transfer for large molecular weight species, our synthesized fluorosurfactants also demonstrate promising biocompatibility, an essential feature for potential applications in biological systems[14]. We assessed biocompatibility of the synthesized fluorosurfactants by encapsulating yeast cells within microdroplets stabilized by the fluorosurfactants, along with growth medium (YPD broth; see Methods section for details). The encapsulated cells were cultured for 24 h (see Methods section for experimental details). Notably, the yeast cells-maintained mobility within the W/O$_F$ microdroplets stabilized by PFPE(H)-COOCOOC$_2$H$_5$, PFPE(H)$_2$-ED$_{900}$ and PFPE(H)$_2$-PEG$_{600}$ (Fig. 2e; Supplementary Fig. 12a-b). We observed no adherence of the yeast cells to the microdroplets interface (Supplementary Movie 1), and their division and proliferation followed the typical pattern of bud formation and gradual bud growth, as previous reported[28]. These findings suggest that our synthesized fluorosurfactants hold considerable potential for applications in 3D cell culturing and bioengineering materials fabrication[4,29].

## 2.3 Tunable functionalities of the synthesized fluorosurfactants

### 2.3.1 Selective enrichment of fluorophores

An intriguing observation was the emergence of nanodroplets emerged from parent W/O$_F$ microdroplets upon the application of 5.0 mM PFPE(H)-COOCOOC$_2$H$_5$. Notably, nanodroplets did not form during the initial generation of the parent W/O$_F$ microdroplets; instead, they appeared immediately after the parent W/O$_F$ microdroplets had formed, regardless of whether fluorophores were encapsulated inside (Fig. 3a). In contrast, significantly fewer nanodroplets were observed when PFPE(H)$_2$-ED$_{900}$ or PFPE(H)$_2$-PEG$_{600}$ was applied at the same concentration (Supplementary Fig. 13a-b). This observation was further confirmed by time-lapse experiment, as shown in Supplementary Movies 2 and 3, corresponding to W/O$_F$ microdroplets stabilized with PFPE(H)-COOCOOC$_2$H$_5$ and PFPE(H)$_2$-ED$_{900}$, respectively. Based on recent studies on nonionic surfactant like Span 80[30,31], we elucidate that hydrogen bonds formed between PFPE(H)-COOCOOC$_2$H$_5$ RMs and H$_2$O, driven by the high electronegativity of the mixed anhydride in PFPE(H)-COOCOOC$_2$H$_5$, facilitated nanodroplets formation through spontaneous emulsification[32]. Moreover, we found that rhodamine 110 chloride was selectively partitioned into nanodroplets while resorufin sodium salt remained in the parent W/O$_F$ microdroplets (Fig. 3b; Supplementary Fig. 14). We postulate that this selectivity is governed by the hydrophilicity/hydrophobicity of the fluorophores[33], which warrants further investigation in future studies. We remark that such selective enrichment realized by PFPE(H)-COOCOOC$_2$H$_5$-stabilized W/O$_F$ microdroplets demonstrates considerable potential for enrichment in microsystems, such as selectively concentrating peptides[34].

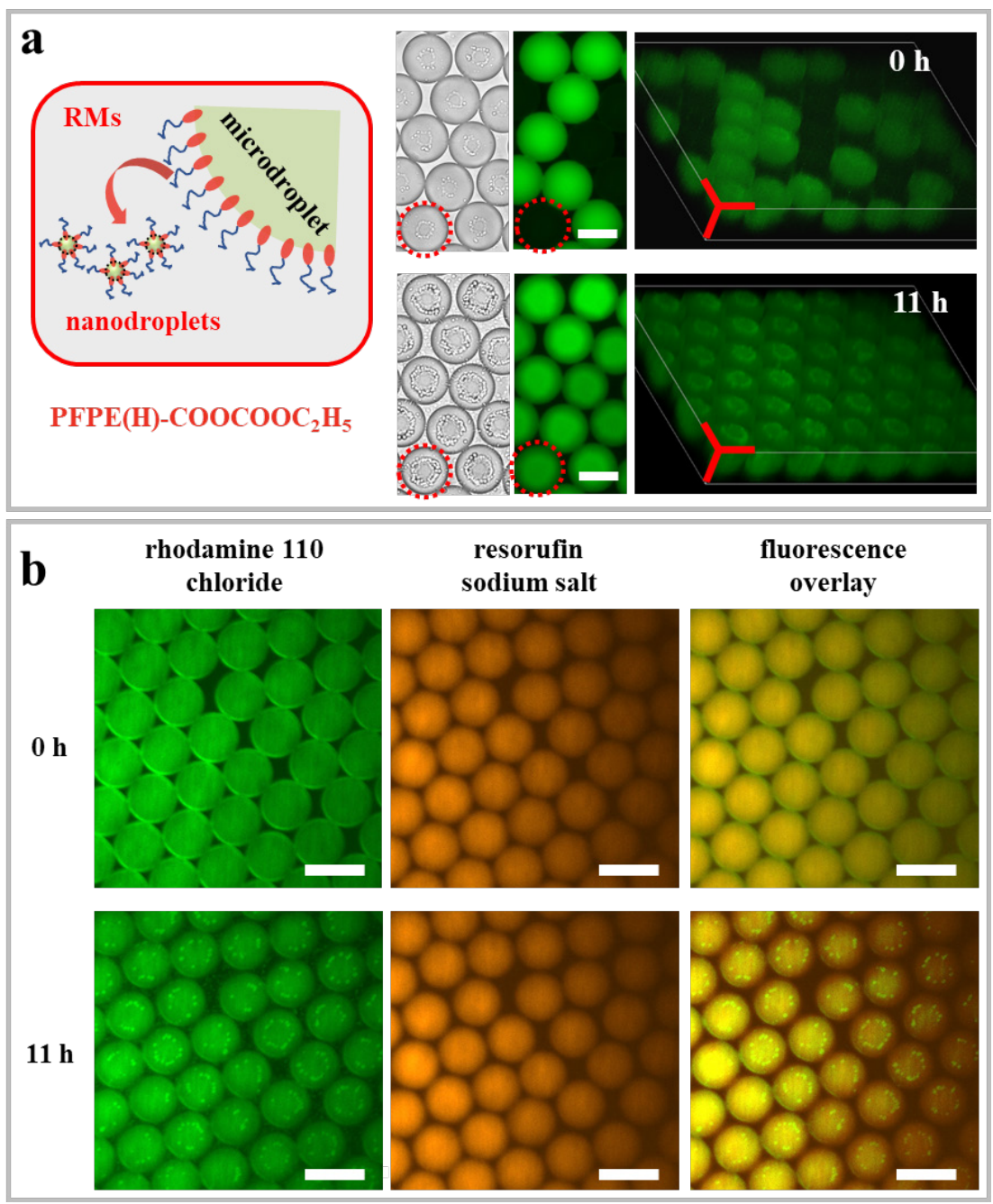

**Figure 3. Selective enrichment of fluorophores.** a) Schematic illustration of nanodroplets formation *via* spontaneous emulsification, accompanied by two sets of representative time-lapse images of microdroplets stabilized by PFPE(H)-COOCOOC$_2$H$_5$. Bright-field (left) and corresponding fluorescence micrographs (middle), along with 3D reconstructions from confocal microscopy images (right), of bi-microdroplets, both immediately after preparation and after 11 h. Typical microdroplets containing nanodroplets can be clearly visualized as marked in red dashed circles for clarity. b) Time-lapse fluorescence images showing selective partitioning of rhodamine 110 chloride from PFPE(H)-COOCOOC$_2$H$_5$ stabilized parent W/O$_F$ microdroplets, containing both rhodamine 110 chloride and resorufin sodium salt, into nanodroplets. Note that fluorescence intensity was extracted from the raw (grayscale) fluorescence micrographs, with false colors applied to enhance the visibility of the distribution and intensity. Scale bars: 100 μm.

## 2.3.2 Engineering morphology-controlled superstructure through nanoparticle self-assembly in slowly evaporating microdroplets.

Microdroplets are widely used as confinement systems for colloidal particle self-assembly, offering an ideal platform for structuring hierarchical materials with diverse applications[35,36,37,38,39,40,41,42]. We synthesized polyvinylpyrrolidone (PVP) stabilized copper oxides ($Cu_xO$) NPs (Supplementary Fig. 15a-b), which show great potential in catalysis and optoelectronics[43, 44]. The synthesized $Cu_xO$ NPs were then assembled into morphology-controlled superstructures, facilitated by the slowly evaporation of a 1:1 water-ethanol (v/v) mixture-in-fluorinated oil ($M/O_F$) microdroplets stabilized by the fluorosurfactants (Fig. 4). In contrast to the coalescence observed in microdroplets stabilized by the commercial fluorosurfactant PFPE(H)-COOH (Supplementary Fig. 16a-b), microdroplets containing $Cu_xO$ NPs remained highly stable when stabilized by the synthesized fluorosurfactants (Fig. 4a-c). As the inner-phase solvent evaporated under ambient conditions, the $Cu_xO$ NPs self-assembled into hierarchical superstructures, displaying red hue in optical microscopy images captured in reflective mode (Fig. 4d-e).

    To understand the role of fluorosurfactants in the formation of the superstructures, we characterized the self-assembled structures using scanning electron microscopy (SEM). Notably, two distinct superstructures were observed, depending on the fluorosurfactants applied during the self-assembly process. Specifically, when stabilized with PFPE(H)-COOCOOC$_2$H$_5$, the resulting superstructure resembled capsule-like colloidosomes with wrinkled surface ($Cu_xO$ NP-CSs; Figure 4f; Supplementary Fig. 17). Additionally, we observed a fiber-like organic matrix, which most likely corresponds to the PFPE(H)-COOCOOC$_2$H$_5$ fluorosurfactants, connecting the $Cu_xO$ NPs (right panel of Fig. 4f). In contrast, focused ion beam SEM (FIB-SEM)

imaging revealed that $Cu_xO$ NPs packed into spherical supraparticles ($Cu_xO$ NP-SPs) when $PFPE(H)_2$-$ED_{900}$ was applied, with no fiber-like structures present (Fig. 4g and Supplementary Fig. 18). This structural variation highlights the role of fluorosurfactants in directing the self-assembly of $Cu_xO$ NPs from slowly evaporating droplets, as elaborated future below.

To investigate the mechanisms underlying the formation of these two distinct superstructures, we applied FT-IR and X-ray photoelectron spectroscopy (XPS) techniques to study the colloidal self-assemblies. The C-F bond vibration peak at ~1254 cm$^{-1}$ from the synthesized fluorosurfactants was clearly observed on both $Cu_xO$ NP-CSs and $Cu_xO$ NP-SPs (Supplementary Fig. 19a), indicating the adsorption of the fluorosurfactants on both self-assemblies[18]. The characteristic peak at 1787 cm$^{-1}$ corresponding to the mixed anhydride C=O stretch in PFPE(H)-COOCOOC$_2$H$_5$ disappeared in $Cu_xO$ NP-CSs, suggesting alternation of the C=O bond environment, likely due to the formation of new bonds between PFPE(H)-COOCOOC$_2$H$_5$ and $Cu_xO$ NPs. In contrast, the amide C=O stretch at 1695 cm$^{-1}$ in $PFPE(H)_2$-$ED_{900}$ remained almost unchanged in $Cu_xO$ NP-SPs, indicating minimal or no chemical bonding between $Cu_xO$ NPs and $PFPE(H)_2$-$ED_{900}$.

The F 1s XPS signal confirms the presence of fluorosurfactants on both $Cu_xO$ NP-CSs and $Cu_xO$ NP-SPs, aligning well with the FT-IR measurements (Supplementary Fig. 19b). The characteristic XPS signal corresponding to $Cu^{2+}$ and $Cu^+$ suggests mixture of $Cu_2O$ and $CuO$ in both isolated/individual NPs and self-assembled superstructures (Fig. 4h). Compared to the isolated/individual $Cu_xO$ NPs, the binding energies of Cu $2p_{3/2}$ and Cu $2p_{1/2}$ in $Cu_xO$ NP-CSs shifted to higher values by approximately 0.9 eV and 0.73 eV, respectively. In contrast, the shifts in the binding energies of Cu $2p_{3/2}$ and Cu $2p_{1/2}$ in $Cu_xO$ NP-SPs were less pronounced, with Cu $2p_{3/2}$ and Cu $2p_{1/2}$ increasing by approximately 0.22 eV and 0.28 eV, respectively (Fig. 4h; Supplementary Table 2). This can be ascribed to stronger interaction between the mixed anhydride group of PFPE(H)-COOCOOC$_2$H$_5$ with copper oxides compared to that between the amide group of $PFPE(H)_2$-$ED_{900}$ with copper oxides. In the XPS spectra of $O_{1s}$, alongside the signal from lattice oxygen ($O_{latt}$) bonded to the $Cu^{2+}$/ $Cu^+$ cations, we observed a peak corresponding to oxygen that is chemically adsorbed ($O_{ads}$) on the surface of the copper oxides NPs. This chemisorbed oxygen exhibits a higher binding energy (Fig. 4i) [45]. We further identified the -$CF_3$ and -$CF_2$- groups in the $Cu_xO$ NP-CSs and $Cu_xO$ NP-SPs, in the C 1s of XPS spectra, indicating the presence of the fluorosurfactants (Fig. 4j). These results unambiguously proved coordination bonding with different strength between the fluorosurfactants (*i.e.* C=O group in the mixed anhydride of the PFPE(H)-COOCOOC$_2$H$_5$ or C=O group in the amide bond of the PFPE(H)$_2$-ED$_{900}$ in this study) with copper cations[46, 47].

We remark that the confined self-assembly process can be considered an equilibrium, as the drying was slow under ambient conditions. Therefore, kinetic effects did should *not* have played a role during the superstructure formation. We propose that the coordination bonding between the PFPE(H)-COOCOOC$_2$H$_5$ through the C=O group in the mixed anhydrides and the copper ions, recruits the $Cu_xO$ NPs to the interface of the M/O$_F$ microdroplets, creating void spaces in the M/O$_F$ microdroplets during the self-assembly process[48]. As a result, capsule-like $Cu_xO$ NP-CSs were formed after the evaporation of the inner phase, illustrated in Fig. 4k, which is reminiscence of recent work reported by Fujiwara *et al.*[49] In contrast, for the M/O$_F$ microdroplets stabilized by PFPE(H)$_2$-ED$_{900}$, the weaker interactions between the $Cu_xO$ NPs and PFPE(H)$_2$-ED$_{900}$ led to a more homogenous distribution of $Cu_xO$ NPs inside the droplets, ultimately resulting in the formation of solid $Cu_xO$ NP-SPs (Fig. 4l). We highlight that our synthesized fluorosurfactants, with their distinct functional groups, provide a versatile means of modulating self-assembled superstructures, enabling precise control over their catalytic and optoelectronic properties.

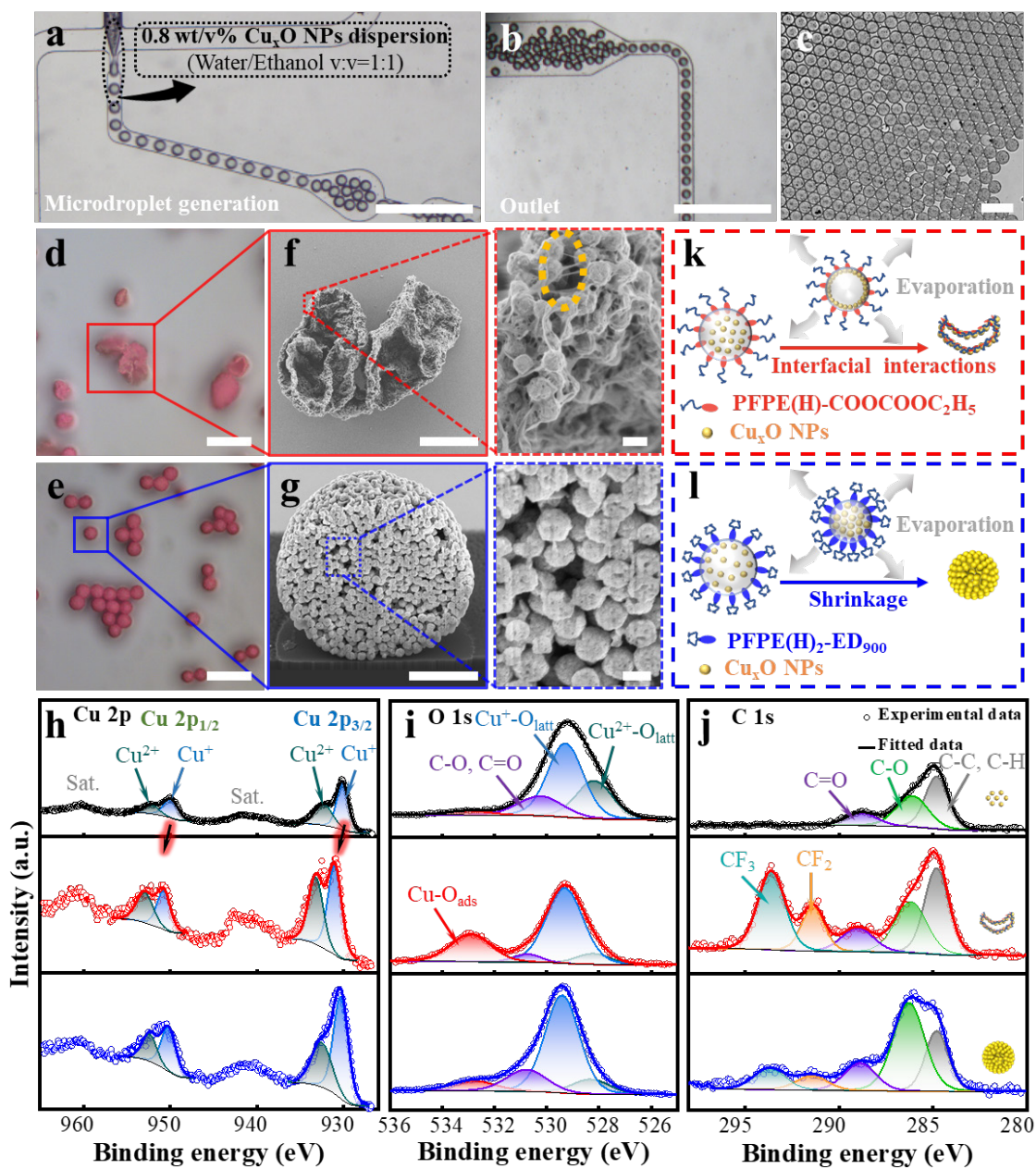

**Figure 4. Structure-controlled self-assembly of $Cu_xO$ NPs in slowly drying M/O$_F$ microdroplets.** a-c) Optical microscopy image of a mixture of water/ethanol in fluorinated oil (M/O$_F$) microdroplets formation using a flow-focus generator (a), outlet of the chip (b), and generated monodisperse M/O$_F$ microdroplets (c). d-e) Optical microscopy image of $Cu_xO$ NP-colloidosomes ($Cu_xO$ NP-CSs) and $Cu_xO$ NP-supraparticles ($Cu_xO$ NP-SPs). f,g) SEM images of $Cu_xO$ NP-CSs and $Cu_xO$ NP-SPs. Right panel of f: a zoomed-in view of the wrinkled surface of the $Cu_xO$ NP-CSs, where a yellow ellipse denotes possible presence of PFPE(H)-COOCOOC$_2$H$_5$ inside $Cu_xO$ NP-CSs. Right panel of g: a zoomed-in view of the cross-sectional morphology of a $Cu_xO$

NP-SP by focus ion beam (FIB) technique. The Cu 2p (h), O 1s (i), and C 1s (j) XPS spectra of the $Cu_xO$ NPs (top row of panels h,i,j), $Cu_xO$ NP-CSs (middle row of panels h,i,j) and $Cu_xO$ NP-SPs (bottom row of panels h,i,j), respectively; k-l) Schematic illustration of possible formation process of $Cu_xO$ NP-CSs and $Cu_xO$ NP-SPs guided by different fluorosurfactants. Scale bars: 500 μm (a-b), 100 μm (c), 20 μm (d-e), 10 μm (left panel of f), 2 μm (left panel of g), 200 nm (right panels of f and g).

## 3. Conclusion

To conclude, we present a facile and robust strategy for synthesizing fluorosurfactants with tunable functionalities through a two-step reaction. The proposed approach offers distinct advantages compared to the state-of-the art methods, including less time-consuming, versatility, less dependent on critical synthetic environment such as high temperature and vacuum. We demonstrated the high droplets stability enabled by our fluorosurfactants even at elevated temperatures. Additionally, we successfully cultured yeast cell in microdroplets, highlighting the bio-inertness of our synthesized fluorosurfactants. We identified hydrogen bonding between the diblock fluorosurfactant PFPE(H)-COOCOOC$_2$H$_5$ and H$_2$O as the driving force behind nanodroplets formation, enabling selective transfer of fluorophores in a binary mixture, and opening up a new avenue for enrichment in biosystems. Furthermore, by modulating the interactions between the NPs and fluorosurfactants, we constructed colloidal superstructures with distinct morphologies. We envisage that the proposed synthetic route will facilitate the robustness and functionality of fluorosurfactants, and driving advancements in a broad range of droplets-based applications.

## 4. Methods

**Materials.** All chemicals were used as provided without further purification unless noted otherwise. DuPont Krytox 157-FSH (carboxylated perfluoropolyether, PFPE(H)-COOH, $M_w$ = 7000~7500 g/mol) was purchased from Miller-Stephenson Chemical Co.

Inc. (Danbury, CT, USA). 3M Novec™ HFE 7100 and 3M Novec™ HFE 7500 were purchased from Miller- 3M (St. Paul, MN, USA). Jeffamine ED$_{900}$, triethylamine (TEA, ≥99.5%), N-methylmorpholine (NMM, 99%), benzotrifluoride (BTF), rhodamine 110 chloride, resorufin sodium salt, FITC-CM-Dextran (150 KDa), CuSO$_4$·5H$_2$O (≥ 98.0%), ethylene glycol (≥ 99.0%), YPD Broth and OptiPrep™ (D1556) were purchased from Sigma-Aldrich. PEG-diamine with a molecular weight of 600 g·mol$^{-1}$ (PEG$_{600}$) was obtained from Shanghai Yare Biotech, Co., Ltd (Shanghai, China). Deionized (DI) water (18.2 MΩ·cm) was prepared using a Milli-Q Plus water purification system (Milli-Q Plus water purification, Sichuan Wortel Water Treatment Equipment Co. Ltd, Chengdu, China). All other chemicals were analytical grades and received from Guangzhou chemical reagent Co., Ltd. (Guangzhou, China).

**Synthesis of diblock- and triblock fluorosurfactants.** The fluorosurfactants reported in this work were synthesized using a two-step process involving mixed anhydride and amidation reactions[50]. In a typical synthesis of the diblock fluorosurfactant (PFPE(H)-COOCOOC$_2$H$_5$), 157-FSH (1.0 mmol) and NMM (1.2 mmol) in 3M Novec™ HFE 7100 (10.0 mL) were mixed in a round-bottom flask with stirring, followed by the addition of ethyl chloroformate (1.2 mmol) at 0 °C for 0.5 h. The triblock fluorosurfactants were synthesized by adding the above mixture to a mixture solution containing Jeffamine ED$_{900}$ or PEG$_{600}$-diamine (0.5 mmol) and triethylamine (1.2 mmol) in BTF (6.0 mL) with stirring at 0 °C for 10 mins. The reaction was kept at RT (25 °C) with stirring for 0.5 h. The synthesized diblock- and triblock fluorosurfactants were purified by washing with ethanol 5 times, and dried under vacuum at 60 °C for 24 h to yield a transparent, viscous liquid. The yield of fluorosurfactants was calculated *via* weighing the synthesized compounds.

**Fourier-transform infrared spectroscopy (FT-IR) analysis.** FT-IR spectra were recorded on a Vertex 70 FT-IR spectrometer (Bruker, Germany) in the transmittance mode. The wavenumber range was 400-4000 cm$^{-1}$, and the resolution was 2.0 cm$^{-1}$.

**$^{19}$F nuclear magnetic resonance (NMR) characterization.** $^{19}$F NMR spectra were acquired using a mixture of benzene-D$_6$ and hexafluorobenzene a 1:6 (v/v) ratio as a solvent *via* ADVANCE NEO 600 MHz NMR system (Bruker, Germany).

**Interfacial tension measurements.** Interfacial tension γ between water and Novec$^{TM}$ HFE 7500 oil was measured by Pendant drop technique using an OCA 15 Pro instrument (Dataphysics, Germany) at RT. The water droplet was injected with 0.1 µL/min and suspended vertically using a syringe needle into a glass cuvette containing Novec$^{TM}$ HFE 7500 oil dissolving fluorosurfactants. The measurements were repeated three times for each sample. The interfacial tension γ was calculated by fitting the droplets shape to the Laplace-Young equation. The resulting data were fit to curves of the form $y(x) = a \cdot \exp(-\beta \cdot x) + \eta$, where the coefficients $a$, $\beta$ and $\eta$ were extracted using a weighted algorithm implemented in MATLAB[51].

**Polydimethylsiloxane (PDMS) droplet generator fabrication.** The PDMS droplet generators with either a flow-focusing geometry (Supplementary Fig. 4a) or a double T junction geometry (Supplementary Fig. 4b) were fabricated using the standard soft lithography technique, as described in our previous article[52]. The mold was created from a silicon wafer with the SU-3050 photoresist to form the designed channel. The based and curing agent was mixed under stirring with a mass ratio of 10:1, followed by degassed in a vacuum chamber until interior bubbles disappeared. Then, PDMS in liquid form was prepared by mixing degassed based and curing agent to duplicate the pattern on the mold. Both PDMS slide and glass cover slide were treated in a plasma cleaner (PDC-002, Mycro Technologies Co., Ltd, Beijing, China) at a power of 10 W for 60 s, then bonded face-to-face to assemble a microfluidic chip. Finally, hydrophobic microchannel was obtained by injecting the Aquapel® (PPG Industries, PA, USA) into the chip, then flushed with air and heated at 120 ºC for 0.5 h.

**Stability of W/O$_F$ microdroplets.** To test the stability and fluorophores retention at RT, W/O$_F$ microdroplets containing fluorophores were generated using a PDMS generator with a flow-focusing geometry (Supplementary Fig. 5a-b). Two syringe pumps were used to inject continuous phase containing Novec$^{TM}$ HFE 7500 and 5.0 mM fluorosurfactants, and dispersed phase, separately. The flow rate for the continuous phase and disperse phase was set to 5.0 µL/min and 1.0 µL/min. The generated microdroplets were then incubated in a home-made poly (methyl methacrylate) (PMMA) cell for five days (Supplementary Fig. 5c-d).

To investigate thermal stability of the microdroplets stabilized by the synthesized fluorosurfactants, the microdroplets were subjected to 65 ºC for 2.0 h, followed by recovering to 25 ºC (RT). The size of the thermal-treated microdroplets were then measured at RT by optical microscope. Moreover, we applied thermal cycling treatment on the microdroplets to test their stability, which is crucial for various applications such as polymerase chain reaction (PCR)[18]. The thermal cycling program began by ramping the temperature from 25 ºC to 65 ºC at a rate of 2.0 ºC/s, holding at 65 ºC for 40 s. The temperature was then raised from 65 ºC to 95 ºC, maintained at 95 ºC for 15 s, then lowered back to 65 ºC at 2.0 ºC/s and held for another 15 s. This was followed by 34 cycles oscillating the temperature between 65 ºC and 95 ºC. Finally, the system was cooled to RT, after which the microdroplet size was measured.

To further study the inter-droplet mass transfer, bi-W/O$_F$ microdroplets, containing fluorescent microdroplets and blank microdroplets (without fluorophores), were generated by a PDMS generator with a double T junction geometry. Three syringe pumps were employed to inject two disperse phases and continuous phase, consisting of Novec$^{TM}$ HFE 7500 with 5.0 mM fluorosurfactants, into the double T-junction PDMS chip. The flow rate was set to 6.0 μL/min for continuous phases, and 0.6 μL/min for disperse phase with pure DI H$_2$O, and 0.5 μL/min for fluorophore disperse phase, respectively. The generated microdroplets were collected for incubation in a home-made PMMA cell for five days.

**Yeast cells culture.** Yeast cells (0.1 g, Angel yeast Co., Ltd., China) were dispersed in YPD broth medium (10 mL, 50 g/L) at 30 ºC for 15 h. The yeast cell suspension was then diluted with YPD broth medium with a concentration of 0.5 mg/mL. To prepare stable emulsion droplets containing yeast cells, the disperse phase here was yeast cells in YPD broth medium mixed with 15% OptiPrep™ (D1556). Novec$^{TM}$ HFE 7500 oil containing synthesized fluorosurfactants at the concentration of 5.0 mM was applied as continuous phase. Note that OptiPrep™ is a biocompatible and non-invasive reagent that increases the density of the aqueous phase, thereby preventing cell aggregation[53]. The W/O$_F$ microdroplets containing yeast cells were generated using a flow-focus droplet generator. Two syringe pumps were applied to regulate the flow of two phases.

We set the flow rates of continuous and phase and dispersed phase at 5.0 μL/min and 1.5 μL/min, respectively. The generated W/O$_F$ microdroplets were collected within a PMMA cell, which were subsequently sealed for continuous observation and culturing under ambient conditions.

**Microscopy image acquisition and data processing.** The generation process of W/O$_F$ microdroplets was recorded using an inverted optical microscope (Olympus IX2, Tokyo, Japan) equipped with a high-speed camera (Phantom, MIRO MIIO, Vision Research Inc., Wayne County, NC, USA), operated in bright-field mode. Fluorescence images were taken by a microscope Leica DMi 5000 M camera with a pE300-ultra LED illumination system (CoolLED, Andover, UK) using the green LED light (450 nm center wavelength) or the blue LED light (537 nm center wavelength) combined with a BGR filter cube (dichroic mirror: 510 nm). The 3D rendering was reconstructed by Fiji from a series of fluorescence images recorded with a Nikon Eclipse Ti-U inverted microscope with a Hamamatsu digital camera (ORCA-flash 4.1, ~100 fps) and 100 mW 488 nm and 561 nm excitation laser sources for rhodamine 110 chloride and resorufin sodium salt, respectively. The size distribution of the generated W/O$_F$ microdroplets and its fluorescence intensity were analyzed by customized Matlab code.

**Synthesis of Cu$_x$O NPs.** We synthesized 166 nm Cu$_x$O NPs according to previous protocol with minor modification (Supplementary Fig. 15)[54]. 0.2496 g of CuSO$_4$ and 0.04 g of PVP were dissolved in 50 mL of the ethylene glycol in a 250 mL three-necked flask under an ultrasonic agitation for 30 mins. Next, 25 mL of an aqueous solution containing 0.1 g of NaOH was added to the above solution and stirred for 10 mins to ensure complete mixing. A glucose solution (6.0 g in 25 mL of DI water) was then added dropwise to the mixed solutions over 15 mins while stirring slowly. The obtained mixture was transferred to a water bath and kept at 80 ºC for 2.0 h. Finally, the resulting orange sediment was purified with a centrifugation force of 6, 940 × g, and washed three times using water and ethanol, respectively. The Cu$_x$O NP dispersion was stored in ethanol. XPS results confirmed presence of both Cu$^+$ and Cu$^{2+}$, indicating partially oxidation during sample storage. Consequently, we designated the synthesized NPs as Cu$_x$O NPs.

**Self-assembly of $Cu_xO$ NPs into colloidosomes (CSs) and supraparticles (SPs).** To enable the self-assembly of $Cu_xO$ NPs from drying microdroplets, we dispersed the $Cu_xO$ NPs in a 1:1 water-ethanol (v/v) mixture with a concentration of 0.8 wt/v%. The mixture-in-fluorinated oil (M/$O_F$) microdroplets were generated using PDMS generator with a flow-focus geometry, and stabilized by fluorosurfactant at concentration of 5.0 mM. Two syringe pumps were used to inject the dispersed phase and continuous phase into the flow-focus PDMS chip, with a flow rate of 5.0 µL/min for continuous phases and 2.5 µL/min for dispersed phase, respectively. The M/$O_F$ microdroplets were subsequently collected in glass vials under ambient conditions, allowing them to dry over approximately five days at RT. The obtained $Cu_xO$ NP-CSs and $Cu_xO$ NP-SPs were purified by washing with Novec[TM] HFE 7100 oil four times, and were stored in Novec[TM] HFE 7100 oil for further characterization.

**X-ray photoelectron spectroscopy (XPS) measurements.** The composition of the synthesized $Cu_xO$ NPs and $Cu_xO$ NP-assemblies were analyzed using X-ray photoelectron spectroscopy (XPS, ESCALAB 250, USA). In-depth XPS profiles were obtained by argon-ion beam etching with a rate of 0.06 nm/s. All XPS spectra were calibrated to the carbon peak (C 1s, 284.8 eV). The XPS peaks were fitted using CasaXPS software, ensuring that the full widths at the half-maximum (FWHM) of the main peaks were kept below 2.0 eV, with a fixed Lorentzian/Gaussian ratio of 20%.

**Scanning electron microscopy (SEM) and focused-ion beam SEM (FIB-SEM) sample preparation and measurement.** To prepare the sample for SEM, 10.0 µL of $Cu_xO$ NP-CSs and $Cu_xO$ NP-SPs were drop-casted on a silicon wafer, and subsequently were dried under ambient condition. The structure of $Cu_xO$ NPs and $Cu_xO$ NP-assemblies were characterized by GeminiSEM 500 (Carl Zeiss Microscopy GmnH) with an accelerating voltage of 0.75 kV in secondary electron (SE) imaging mode. To further analyze the interior morphology of $Cu_xO$ NP-SPs, $Cu_xO$ NP-SPs was sectioned by focused ion beam (FIB, Thermo scientific Helios 5 UX) using a Ga ion source. The milled cross-section was imaged in secondary electron mode at an accelerating voltage of 5.0 kV.

## Data availability

The authors declare that all the relevant data are available within the paper and its Supplementary Information file or from the corresponding authors upon reasonable request.

## Acknowledgements

We acknowledge financial support from the National Natural Science Foundation of China (No. 12374211), and Science and Technology Project of Guangdong Province (No. 2024A1515010687, 2023A1515010935, 2022A1515220144). This work has also been partially supported by PCSIRT Project No. IRT_17R40, the National 111 Project, the MOE International Laboratory for Optical Information Technologies, and China Scholarship Coucil (CSC) (Grant No. 202106750031). The authors acknowledge Daria Bugakova for her help in droplet generator fabrication.

## Author contributions

L. Shui initiated the project. J. Yao designed and synthesized the fluorosurfactants. J. Yao, S. Huang, Y. Deng, Z. Liu conducted the droplet microfluidics experiments. J. Yao, S. Xie and Y. Deng performed the colloidal self-assembly experiments. J. Yao analyzed the fluorescence intensity using customized code implemented in MATLAB develop by L. Carnevale. All authors contributed to discussion and analysis of the experimental results. J. Yao, S. Huang, S. Xie, M. Jin, D. Wang, S. Pud, L. Shui and L. I. Segerink wrote the manuscript.